\documentclass [12pt]{article}

\usepackage{natbib}
\usepackage {graphicx}
\usepackage{epsfig,amsmath,latexsym,amssymb}
\usepackage{bm}

\usepackage{booktabs}
\usepackage{multirow}
\usepackage{float,lscape}
\usepackage{rotating}

\newtheorem{algorithm_new}{Algorithm}[section]
\newtheorem{remark}{Remark}[section]

\begin{document}


\title{\vspace{-2cm} Fast and Simple Method for Pricing Exotic Options using Gauss-Hermite Quadrature on a Cubic Spline Interpolation}

\author{Xiaolin Luo$^{1}$ and Pavel V.~Shevchenko$^{2}$}

\date{\footnotesize{This version 25 November 2014}}

\maketitle

\vspace{-1cm}

\begin{center}
\footnotesize { \textit{$^{1}$ The Commonwealth Scientific and Industrial Research Organisation, Australia \\e-mail: Xiaolin.Luo@csiro.au \\
$^{2}$ The Commonwealth Scientific and Industrial Research
Organisation, Australia\\
e-mail: Pavel.Shevchenko@csiro.au } }
\end{center}
\begin{abstract}
\noindent There is a vast literature on numerical valuation of
exotic options using Monte Carlo, binomial and trinomial trees, and
finite difference methods. When transition density of the underlying
asset or its moments are known in closed form,
 it can be convenient and more efficient to utilize direct
integration methods to calculate the required option price
expectations in a backward time-stepping algorithm. This paper
presents a simple, robust and efficient algorithm that can be
applied for pricing many exotic options by computing the
expectations  using Gauss-Hermite integration quadrature applied on
a cubic spline interpolation.   The algorithm is fully explicit but
does not suffer the inherent instability of the explicit finite
difference counterpart. A `free' bonus of the algorithm is that it
already contains the function for fast and accurate interpolation of
multiple solutions required by many discretely  monitored path
dependent options.  For illustrations, we present examples of
pricing a series of American options with either Bermudan or
continuous exercise features, and a series of exotic path-dependent
 options of target accumulation redemption note (TARN). Results of the new
method are compared with Monte Carlo and finite difference methods,
including
 some of the most advanced or best known finite difference algorithms in the
 literature.  The comparison  shows that, despite its simplicity, the new method can rival with
some of the best finite difference algorithms in accuracy and at the
same time it is significantly faster.  Virtually the same algorithm
can be  applied to price other path-dependent
 financial contracts such as Asian options and variable annuities.

\hspace{1cm}

\end{abstract}

\pagebreak

\section{Introduction}
There is a vast literature on numerical valuation of exotic options
using Monte Carlo, trees and partial differential equation methods.
For text book treatment of this topic, see \citet{wilmott2007paul}
and \citet{hull2009options}; also more specialized books
\citet{glasserman2004monte} for Monte Carlo and
\citet{tavella2000pricing} for finite difference methods. The choice
of method is dictated by the type of option and underlying asset
stochastic process. For example, for pricing American type option
the modeller can use finite difference or tree methods in the case
of one or two underlying assets and the Least-Squares Monte Carlo
(LSMC) method (\citet{longstaff2001}) for higher dimensions; for
pricing basket options without early exercise features the modeller
can use standard Monte Carlo; etc. It is also worth mentioning that
for contracts where the underlying asset process depends on the
contract control variables (e.g. variable annuities with guaranteed
minimum withdrawal benefit where the underlying wealth process is
affected by optimal cash withdrawals that should be found from
backward solution), the underlying process cannot be simulated
forward in time and thus the standard LSMC method cannot be applied.

In the case when transition density of the underlying asset between
time slices or its moments are known in closed form and problem
dimension is low (one or two underlying stochastic variables), often
it can be convenient and more efficient to utilize direct
integration methods to calculate the required option price
expectations in backward time-stepping algorithm. In this paper, we
present a method that relies on computing the expectations in
backward time-stepping through Gauss-Hermite integration quadrature
applied on a  cubic spline interpolation; we use GHQC (Gauss-Hermite
quadrature on Cubic-spline) to denote this method. We show that the
GHQC can be made fully explicit, so it is as fast as an explicit
finite difference algorithm but at the same time it is more accurate
and does not suffer the inherent instability of the latter. This
approach can be applied to numerical valuation of many exotic
options including barrier, Asian and American type options, and
contracts written on the asset with path affected by optimal control
variables such as variable annuity with guaranteed minimum
withdrawal benefit. In general, it can be applied to any option that
can be evaluated using finite difference method if the underlying
asset transition density or its moments are easily evaluated. It is
easy to implement and understand. Also, in general the  accuracy
 of GHQC can rival
 the widely used semi-implicit second order finite difference algorithm, but GHQC is much faster
  because it either requires less
number of time steps or it is faster per time step due to its full
explicitness. Of course the number of time steps depends on time
discretisation required by stochastic process and option contract
details (e.g. barrier or early exercise monitoring frequency).

We first describe GHQC algorithm in details and then illustrate the
use of the algorithm to compute American options and the
path-dependent TARN options. Two series of American options are
considered - one with discrete exercise (so called Bermudan option)
and another one with continuous exercise in time. The results of
GHQC are compared with those by several finite difference algorithms
and LSMC method. The comparison includes some of the most
sophisticated and advanced finite difference algorithms found in the
literature (\citet{Brennan1977}, \citet{Wilmott1995},
\citet{Leisen1996}, \citet{Kwok1997}, \citet{Forsyth2002},
\citet{Nielsen2002}, \citet{Han2003}, \citet{Ikonen2004},
\citet{Borici2005}, and \citet{Tangman2008}), and it
 demonstrates the good accuracy and high efficiency of the algorithm.

It is straightforward to apply the algorithm (with similar benefits)
to barrier options, Asian options, targeted accrual redemption notes
(TARNs) and variables annuities. This is demonstrated by pricing a
series of twelve TARN contracts covering three  ``knockout" types
and four accumulation target levels. The results are compared with
 those of finite difference and Monte Carlo, which again
 illustrates the robustness,  accuracy and efficiency of the GHQC
 algorithm.
  Of course it is expected that
in high dimensions the Monte Carlo method will be more efficient
(standard MC for options without early exercise and LSMC for
American type options). For application of the algorithm for pricing variable annuities with
GMWB and death benefit, see \citet{LuoShevchenko2014gmwb, LuoShevchenko2014gmwdb}.

\section{Model}
Let $S(t)$ denote the value of the option underlying asset that
follows the risk-neutral stochastic process
\begin{equation}\label{stochasticmodel_eq}
dS(t)=\mu(t) S(t) dt+\sigma(t) S(t) dB(t),
\end{equation}
\noindent where $\mu(t)$ is the drift (i.e. it is the risk-free
interest rate $r(t)$ minus dividends if $S(t)$ is equity or
difference between domestic and foreign interest rates if $S(t)$ is
foreign exchange), $\sigma(t)$ is the volatility and $B(t)$ is a
standard Brownian motion. The risk free interest rate $r(t)$ can be
function of time. The drift and volatility can be functions of time
and underlying asset but for illustrative examples we assume that
drift and volatility are functions of time only.

In this paper we do not consider time discretization errors; for
simplicity, hearafter, we assume that model parameters are
piece-wise constant functions of time for discretization
$0=t_0<t_1<\cdots<t_N=T$, where $T$ is contract maturity. Denote
corresponding asset values as $S(t_0),S(t_1),\ldots,S(t_N)$;  and
drift, risk-free interest rate and volatility as
$\mu_1,\ldots,\mu_N$, $r_1,\ldots,r_N$ and
$\sigma_1,\ldots,\sigma_N$ respectively. That is $\mu_1$ is the
drift for time period $(t_0,t_1]$; $\mu_2$ is for $(t_1,t_2]$, etc.
and similar for risk-free interest rate and volatility. To simplify
notation, we also assume that the monitoring frequency specific to
the option contract corresponds to the same time discretization. For
example, barrier monitoring dates or Bermudan exercise dates or
Asian averaging dates are the same as the time discretization. It is
trivial extension if monitoring dates reside on some time slices
only so that there are time steps between the monitoring dates.

Denote the transition density function from $S(t_{n-1})$ to $S(t_n)$
as $p_n(s(t_n)|s(t_{n-1}))$, which is just a lognormal density in
the case of process (\ref{stochasticmodel_eq}) with solution

\begin{equation}\label{eq_asset}
S(t_n)=S(t_{n-1})e^{(\mu_n-\frac{1}{2}\sigma_n^2)dt_n+\sigma_n
\sqrt{dt_n} z_n}, \;\; n=1,2,\ldots,N,
\end{equation}
where $dt_n=t_n-t_{n-1}$ and $z_1,\ldots,z_N$ are independent and
identically distributed random variables from the standard Normal
distribution. That is, distribution of $\ln S(t_n)$ conditional on
$\ln S(t_{n-1})$ is Normal with the mean $\ln S(t_{n-1}) +
(\mu_n-\frac{1}{2}\sigma_n^2)dt_n$ and standard deviation $\sigma_n
\sqrt{dt_n}$.

In general, the today's fair price of the option can be calculated
as expectation of discounted option payoff with respect to the risk
neutral process (\ref{stochasticmodel_eq}), given information today
at $t_0$, and typically can be found via backward time stepping by
calculating option price $Q_n(S(t_n))$ at each $t_n$ that requires
evaluation of conditional expectations (conditional on information
at $t_{n-1}$)
\begin{equation}\label{eq_expS}
\widetilde{Q}_{n-1}(S(t_{n-1}))=\mathrm{E}_{t_{n-1}}\left[e^{-r_n
dt_n} {Q_n}\left (S(t_n)\right )\right],
\end{equation}
and applying some jump or early exercise condition specific to the
option. If no condition is applied then $\widetilde{Q}_{n-1}$ equals
${Q}_{n-1}$.  Expectation (\ref{eq_expS}) can be written explicitly
as the following integral
\begin{equation}\label{eq_intS}
\widetilde{Q}_{n-1}(S(t_{n-1}))=\int_0^{+\infty} e^{-r_n dt_n} p_n
\left (s|S(t_{n-1})\right ) Q_n(s) ds.
\end{equation}

The objective of the GHQC algorithm is to evaluate these integrals
efficiently. These integrations occur in many path-dependent
options. Below we explicitly show this for Bermudan, barrier, Asian,
a target accumulation redemption note (TARN) and variable annuity
with Guaranteed Minimum Withdraw Benefit (GMWB).

\subsection{Bermudan option} Consider Bermudan option with early exercise dates the same
as time discretization $0=t_0<t_1<\cdots<t_N=T$. Then  the option
value at $t=0$ is calculated recursively backward in time as
\begin{equation}\label{Bermudan_option_eq}
Q_{n-1}(S(t_{n-1}))=\max \left(\widetilde{Q}_{n-1}(S(t_{n-1})), \max
\left(0, \phi \times(S(t_{n-1})-K)\right)\right),
 \end{equation}
 starting from
$$
Q_N(S(T))=\max (0, \phi \times(S(T)-K)),
$$
  where $\widetilde{Q}_{n-1}(S(t_{n-1}))$ is given by (\ref{eq_intS}), $K$ is the strike, $\phi=1$ for
call option and $\phi=-1$ for put option. In the limit of continuous
early exercise ($N\rightarrow \infty$), this option is called
American option.

\subsection{Barrier option} Consider barrier option with piece-wise constant barriers with
time discretization $0=t_0<t_1<\cdots<t_N=T$. Denote the
lower and upper barriers as $L_1,\ldots,L_N$ and $U_1,\ldots,U_N$
respectively, where $L_1$ is the lower barrier for time period
$[t_0,t_1]$; $L_2$ is for $(t_1,t_2]$, etc. and similar for the
upper barrier. Then the knockout barrier option value at $t=0$ is calculated
recursively backward in time as

\begin{eqnarray}
&&Q_{n-1}(S(t_{n-1}))=\widetilde{Q}_{n-1}(S(t_{n-1}))\nonumber\\
&&\quad =\int_{L_n}^{U_n}
e^{-r_n dt_n} p_n \left (s|S(t_{n-1})\right )g_n(S(t_{n-1}),s)
Q_n(s) 1_{(L_n,U_n)}(S(t_{n-1})) ds
\end{eqnarray}
starting from
$$
Q_N(S(T))=\max (0, \phi \times(S(T)-K)),
$$
where $K$ is the strike, $\phi=1$ for knockout call option and $\phi=-1$ for
knockout put option, and $1_\mathcal{A}(x)$ is indicator function equal 1 if $x\in\mathcal{A}$ and $0$ otherwise. $g_n(s_{n-1},s_n)$ is probability of no barrier hit
within $[t_{n-1},t_n]$ conditional on asset taking values $s_{n-1}$
and $s_n$ at $t_{n-1}$ and $t_n$ respectively with $s_{n-1}\in (L_n,U_n)$
and $s_n\in (L_n,U_n)$; it is the so-called
Brownian bridge correction often used in the literature on pricing
barrier options, see e.g. \citet{AndersenBrotherton1996,
Shevchenko2003, Beaglehole1997, Shevchenko2014}. In the case of
discrete barrier monitoring
 (i.e. no barrier during $(t_{n-1},t_n)$), $g_n(s_{n-1},s_n)=1$ if both $s_{n–1}$
and $s_n$ are between the barriers and zero otherwise; in the case of single continuous barrier $B_n$ (either
lower or upper) during $[t_{n-1},t_n]$
\begin{equation}
g_n(s,s^\prime)=1-\exp\left(-2\frac{\ln(s^\prime/B_n)\ln(s/B_n)}{\sigma^2_n
d t_n}\right);
\end{equation}
and there is a closed form solution for  the case of continuous
double barrier within $[t_{n-1},t_n]$
\begin{eqnarray}\label{prob_nobarrierhit_eq}
g_n(s,s^\prime)&=&1-\sum_{m=1}^{\infty}[R_n\left(\alpha_n m
-\gamma_n,x_n\right)+R_n(-\alpha_n m+\beta_n,x_n)]\nonumber\\
&&+\sum_{m=1}^{\infty}[R_n(\alpha_n m,x_n)+R_n(-\alpha_n m,x_n)],
\end{eqnarray}
where
$$
x_n=\ln\frac{s^\prime}{s},\alpha_n=2\ln\frac{U_n}{L_n},\beta_n=2\ln\frac{U_n}{s},
\gamma_n=2\ln\frac{s}{L_n},
R_n(z,x)=\exp\left(-\frac{z(z-2x)}{2\sigma^2_n \delta t_n}\right).
$$
If either $s$ or $s^\prime$ breaches the barrier condition, then we get set $g(s,s^\prime)=0$.
Typically only a few terms in summations in
(\ref{prob_nobarrierhit_eq}) are required to achieve good accuracy
for option price estimator.

\subsection{Asian option} Consider a discretely monitored Asian
option with the standard arithmetic average defined over the
monitoring times $t_1,\ldots,t_n$ as
$$A(t_n)=\frac{1}{n}\sum_{i=1}^n S(t_i).$$

Between monitoring times, i.e. for  $t\in (t_{n-1},t_n)$, the
arithmetic average is
 $A(t)=A(t_{n-1})$ and the option value $Q(S(t), A(t))$ can be evaluated as
\begin{equation}\label{eq_intS2}
Q(S(t), A(t))={\widetilde Q}(S(t),A(t))=\int_0^{+\infty} e^{-r_n
(t_n-t)} p_n \left (s|S(t)\right ) Q_{n}(s(t_n^-), A(t_n^-)) ds,
\end{equation}
where $t_n^-$ denotes the time immediately {\it before} the
monitoring time $t_n$. Let $t_{n-1}^+$ denote the time  immediately
{\it after} the monitoring time $t_{n-1}$, then
  the option value $Q_{n-1}(S(t_{n-1}^+), A(t_{n-1}^+))$ can be calculated from  (\ref{eq_intS2}) by letting $t=t_{n-1}^+$ .
 The option value immediately before  $t=t_{n-1}$,
 $Q_{n-1}(S(t_{n-1}^-), A(t_{n-1}^-))$, can then be obtained through a special jump
 condition reflecting the continuity of option value and the finite change in the arithmetic
 average across from $t_{n-1}^-$ to $t_{n-1}^+$:
\begin{equation}\label{eq_jumpAsian}
Q_{n-1}(S(t_{n-1}^-),A(t_{n-1}^-))={\widetilde
Q}_{n-1}\left(S(t_{n-1}^-),A(t_{n-1}^-)+\frac{S(t_{n-1}^-)
-A(t_{n-1}^-)}{n-1} \right ),
\end{equation}
where $A(t_{n-1}^-)=A(t_{n-2})= \frac{1}{n-2}\sum_{i=1}^{n-2}
S(t_i)$ and of course $S(t_{n-1}^-)=S(t_{n-1}^+)=S(t_{n-1})$.
Repeatedly applying (\ref{eq_intS2}) and (\ref{eq_jumpAsian})
backwards gives us the option value at $t=0$, starting from
$$
Q_N(S(T), A(T))=\max (0, \phi\times(S(T)-A(T)).
$$

Note that formula (\ref{eq_intS2}) is for a given fixed value of
$A$, while the jump condition (\ref{eq_jumpAsian}) implies that
across any monitoring
 time the average $A$ jumps to arbitrarily different values
 depending on the values of $A$ and the asset value $S$, which means
 in principle we have to track multiple solutions corresponding to
  all possible values of $A$. Numerically, this can be done by
  discretizing the average space  $A$, similar to discretizing the asset space $S$.
  Interpolating multiple solutions of different values of $A$
  enables
  us to apply the jump condition at all monitoring times.

\subsection{TARN}\label{subsec_TARN} Consider TARN contract that
provides a capped sum of payment (target cap) over a period with the
possibility of early termination.  Let $U$ be the target accrual
level and $A_n=\sum_{i=1}^n \max (0, \phi\times(S(t_i)-K)))$ is the
accumulated amount on the fixing date $t_n$. Later we will show
 numerical results of pricing TARN contracts with three different ``knockout" types used in practice:
\begin{itemize}
\item Full gain -- when the target is breached on a fixing date $t_n
$, the cash flow payment on that date is allowed. This essentially
permits the breach of the target once, and the total payment may
exceed the target for full gain knockout.

\item No gain -- when the target is breached, the entire payment on
that date is disallowed. The total payment will never reach the
target for no gain knockout.

\item Part gain -- when the target is breached on a fixing date $t_n $,
part of the payment on that date is allowed, such that the target is
met exactly.
\end{itemize}

 In the case of ``full gain" knockout TARN,  the present value (discounted value at $t_0=0$) of
the TARN payoff $P^{(\text{Full})}$ is
$$
P^{(\text{Full})}=\sum_{i=1}^{\widetilde N} e^{-rt_i} \max (0,
\phi\times(S(t_i)-K))),
$$
where  $r=(r_1 dt_1+\cdots+r_i dt_i)/t_i$ and $1 \le {\widetilde N}
\le N$ is the first time the target $U$ is breached by $A_n$. In the
case of ``no gain", the last payment is disallowed, thus the payoff
$P^{(\text {No gain})}$ is
$$
P^{(\text{No gain})} =\sum_{i=1}^{{\widetilde N}-1} e^{-rt_i} \max
(0, \phi\times(S(t_i)-K))),
$$
and in the case of ``part gain", we have payoff $P^{(\text{Part})}$
given as
$$
P^{(\text{Part})} = P^{(\text{No gain})} + e^{-rt_{\widetilde
N}}\times(U-P^{(\text{No gain})}).
$$

The price evolution of TARN between payment dates can also be
expressed by (\ref{eq_intS2}), the same as for the Asian option.
However, the jump condition across a payment date now becomes
\begin{equation}\label{eq_jumpTARN}
Q_{n-1}(S(t_{n-1}^-),A(t_{n-1}^-))={\widetilde
Q}_{n-1}\left(S(t_{n-1}),A(t_{n-1}^+)\right)+\max (0, \phi\times
(S(t_{n-1})-K)),
\end{equation}
where $A(t_{n-1}^+)=A(t_{n-1}^-) +\max (0,
\phi\times(S(t_{n-1})-K))$, and $A(t_{n-1}^-) = A(t_{n-2}^+) < U$.
Obviously, similar to the Asian option, tracking of multiple
solutions corresponding to different accumulated amount $A_n$ is
required and thus interpolation between them is necessary. For
numerical valuation of TARN contracts via Monte Carlo and finite
difference methods, see \citet{Piterbarg2004} and
\citet{luo2014pricing}.

\subsection{GMWB} Guaranteed Minimum Withdraw Benefit is one of the
most popular variable annuity contracts in practice, see e.g.
\citet{dai2008guaranteed} and \citet{LuoShevchenko2014gmwb, LuoShevchenko2014gmwdb}. A GMWB contract promises to return the
entire initial investment through cash withdrawals during the policy
life plus the remaining account balance at maturity, regardless of
the portfolio performance. Assume
 the entire initial premium $W(0)$ is invested in asset $S$, and at each
 withdraw date $t_n$ the amount $\gamma_n$ is withdrawn. Then
the account balance of the guarantee $A(t)$ with $A(0)=W(0)$ evolves
as
\begin{equation}\label{accountbalance_eq}
A(t_n)=A(t_{n}^{-})-\gamma_n=A(t_{n-1})-\gamma_n,\;\; n=1,2,\ldots,N
\end{equation}
with $A(T)=0$, $W(0)=A(0) \ge \gamma_1+\cdots+\gamma_N$ and
$A(t_{n-1})= A(t_{n-1}^+) \ge \sum_{i=n}^N\gamma_{i}$.  The value of
personal variable annuity account $W(t)$ evolves as
\begin{equation}
W(t_n)=\max\left[W(t_{n-1})
e^{(\mu_n-\alpha-\frac{1}{2}\sigma_n^2)dt_n+\sigma_n \sqrt{dt_n}
z_n} - \gamma_n,0\right]\;\; n=1,2,\ldots,N,
\end{equation}
where $dt_n=t_n-t_{n-1}$, $z_n$ are independent and identically
distributed random variables from the standard Normal distribution
and $\alpha$ is the annual fee. If the account balance becomes zero
or negative, then it will stay zero till maturity.

 The numerical algorithm  for GMWB with discrete withdrawals is again very similar to Asian options
described above, at least for the ``static" or passive case where
the withdraw amount $\gamma_n=G$ is a constant specified by the
contract. The evolution of price between withdraw dates expressed by
(\ref{eq_intS2})
  still holds, provided the drift $\mu_n$ is replaced by $\mu_n-\alpha$ to account for the
   continuously charged fee $\alpha$. After the
amount $\gamma_n$ is drawn at $t_n$, the annuity account reduces
from $W(t_n^-)$ to $W(t_n) = \max (W(t_n^-) -\gamma_n,0)$,
 and the jump condition of $Q_n(W,A,t)$ across $t_n$ is given
by

\begin{equation}\label{eqn_jumpGMWB}
Q_{n-1}\left (W(t_n^-),A(t_n^-)\right)={\widetilde
Q}_{n-1}\left(\max(W(t_n^-)-\gamma_n,0), A(t_n^-)-\gamma_n
\right)+\gamma_n.
\end{equation}
Repeatedly applying (\ref{eq_intS2}) and (\ref{eqn_jumpGMWB})
backwards gives us the contract value to $t=0$, starting from the
final condition
$$
Q_N(W(T^-), A(T^-))=\max\left(W(T^-), A(T^-)\right).
$$

 For the dynamic (optimal) withdraw case, the
policyholder may decide to withdraw above or below the contractual
rate to maximize the present value of the total cash flow generated
from holding the GMWB contract, and in such a case a penalty is
applied by the insurer and the net cash received by the policyholder
for each withdraw $\gamma$ becomes $c(\gamma)$ which may be less
than $\gamma$:
\begin{equation}
c(\gamma)=\left\{\begin{array}{ll}
                   \gamma, & \mbox{if}\; 0\le \gamma\le G, \\
                   G+(1-\beta)(\gamma-G), & \mbox{if}\; \gamma>G,
                 \end{array} \right.
\end{equation}
and the jump condition for the optimal withdraw case now can be
written as
\begin{equation}
Q_n\left (W(t_n^-),A(t_n^-)\right)=\max_{0 \leq \gamma_n\leq
A(t_n^-) } \left[{\widetilde Q}_n\left(\max(W(t_n^-)-\gamma_n,0),
A(t_n^-)-\gamma_n \right)+c(\gamma_n)\right ],
\end{equation}
That is, at each withdraw date $t=t_n$ the policyholder `optimally'
withdraws an amount $\gamma_n$ to maximize the option value. The
final condition for the dynamic case is given by
$$
Q_N(W(T^-), A(T^-))=\max\left(W(T^-), c\left(A(T^-)\right)\right).
$$

\section{The GHQC Method}
The few path-dependent options described in the previous section all
require the evaluation of expectation (\ref{eq_expS}).  The most
widely used methods for calculating this expectation are pde
solutions and Monte Carlo simulations, and a possible alternative is
 the direct numerical  integration of (\ref{eq_intS}), a recent
 example of this approach can be found in \citet{gheno2014}.

 Except the American and barrier options,  all the other examples
 described in the last section
require accurate interpolation of multiple solutions. As shown in a
convergence study by \citet{Forsyth2002}, it is possible for a
numerical algorithm of discretely sampled path-dependent option
pricing to be non-convergent (or convergent to an incorrect answer)
if the interpolation scheme is selected inappropriately. Typically
previous studies of numerical pde solution for path-dependent (Asian
or lookback options) used either a linear or a quadratic
interpolation in applying the jump conditions.

Below we propose a simple, robust and efficient algorithm for
integrating
 (\ref{eq_intS}) in the context of path-dependent option pricing, and at the
  same time the proposed
  algorithm also naturally (as a byproduct) provides an accurate and efficient procedure for
 interpolating multiple solutions at little extra computing cost.

 \subsection{Numerical evaluation of the expectation}
Similar to a finite difference scheme, we propose to discretize the
asset domain $(S_{\min}, S_{\max}) $  by $S_{\min} =S_0 < S_1,
\ldots,S_M=S_{\max}$ , where $S_{\min}$ and $S_{\max}$ are the lower
and upper boundary, respectively, both are sufficiently far from the
spot asset value at time zero $S(0)$. A reasonable  choice of such
boundaries could be $S_{\max}=S(0) \exp(5\sigma \sqrt{T})$ and
$S_{\min}=S(0) \exp(-5\sigma \sqrt{T})$ (a better choice will be
given later in (\ref{eq_bd})). The idea (from finite difference
method) is to find option values at all these grid points at all
time slices $0 =t_0< t_1 <\ldots<t_N=T$ through backward time
stepping, starting at maturity $t=t_N=T$, and at each time step we
evaluate the integration (\ref{eq_intS}) for every grid point  by a
high accuracy numerical quadrature.

The option value at $t=t_n$
is known only at grid points $S_m$, $m=0,1,\ldots,M$. In order to
approximate the continuous function $Q_n(S(t_n))$ from the values at
the discrete grid points, and perform the required integration, we
propose to
 use the cubic spline interpolation
 which is smooth in the first derivative and continuous in the second derivative (\citet{Pres92}).
 The error of cubic
spline is $O(h^4)$, where $h$ is the size for the spacing of the
interpolating variable, assuming a uniform spacing. Given any
arbitrary tabulated function $Q(x_j),\;j=0,\ldots,M$, the value of
$Q(x), \; x_j < x < x_{j+1}$, can be approximated by the cubic
spline interpolation
\begin{equation}\label{eq_g}
Q(x)\approx AQ(x_j)+BQ(x_{j+1})+CQ''(x_j)+DQ''(x_{j+1}),
 \end{equation}
where
$$A=\frac{x_{j+1}-x}{x_{j+1}-x_j},\;\;B=1-A,$$
$$C=(A^3-A)(x_{j+1}-x_j)^2/6, \;\; D=(B^3-B)(x_{j+1}-x_j)^2/6.$$

From a continuity condition, the second derivatives are obtained by
solving the following tri-diagonal system of linear equations
(\citet{Pres92})
\begin{eqnarray}\label{eq_tri}
&&\frac{d x_j}{6}Q''(x_{j-1})+ \frac{ x_{j+1}-x_{j-1}}{3}Q''(x_{j})+
\frac{dx_{j+1}}{6}Q''(x_{j+1})\nonumber \\
&&\quad\quad\quad\quad\quad\quad\quad
=\frac{Q(x_{j+1})-Q(x_{j})}{dx_{j+1}} -
\frac{Q(x_{j})-Q(x_{j-1})}{dx_j},
\end{eqnarray}
where $\;j=1,\ldots,M-1$,$dx_j=x_j-x_{j-1}$, $dx_{j+1}=x_{j+1}-x_j$.
 For boundary conditions we can set $Q''(x_0)=Q''(x_M)=0$ (natural boundary condition), which is consistent
  with the boundary condition of zero second-derivatives for option values at far boundaries.
Other boundary conditions are possible, depending on the option
specifics.
  For a fixed grid, the tri-diagonal matrix can be inverted once
and at each time
 step only the back-substitution in the cubic spline procedure is
 required. Cubic spline functions are available in most numerical
 packages and are very easy to use.

For the lognormal stock process, the distribution of $S(t_n)$ given
$S(t_{n-1})$  is a lognormal distribution corresponding to solution
 (\ref{eq_asset}). A convenient and common practice is to work
with  $\ln(S(t_n))$ so that the corresponding conditional density is
Normal distribution with the mean $\ln S(t_{n-1}) +
(\mu_n-\frac{1}{2}\sigma_n^2)dt_n$ and standard deviation $\sigma_n
\sqrt{dt_n}$.  In order to make use of the highly efficient
Gauss-Hermite numerical quadrature for integration over an infinite
domain, for each time slice, we introduce a new variable
   \begin{equation}\label{eq_y}
Y(t_n)=\frac{\ln\left(S(t_n)/S(t_{n-1})\right)-\nu_n}{\tau_n},
 \end{equation}
where $\nu_n=(\mu_n-\frac{1}{2}\sigma_n^2 )dt_n$ and
$\tau_n=\sigma_n \sqrt{dt_n}$, and denote the option price function
$Q_n(s)$ after this transformation as $Q_n^{(y)}(y)$. By changing
variable from $S(t_n)$ to $Y(t_n)$ the integration (\ref{eq_intS})
becomes

 \begin{equation}\label{eq_intY}
\widetilde{Q}_{n-1}(S(t_{n-1}))=\frac{ e^{-r_ndt_n}}{\sqrt{2\pi}}
\int_{-\infty}^{+\infty} e^{-\frac{1}{2}y^2} Q_n^{(y)} (y) dy.
\end{equation}
For such  integrals the Gauss-Hermite integration quadrature is well
known to be very efficient (\citet{Pres92}). For an arbitrary
function $f(x)$, the Gauss-Hermite quadrature is
\begin{equation}\label{eq_GHQ}
\int_{-\infty}^{+\infty}e^{-x^2}f(x)dx \approx \sum_{j=1}^q
\lambda_j^{(q)} f(\xi_j^{(q)}),
\end{equation}
where  $q$ is the order of the Hermite polynomial, $\xi_j^{(q)}$ are
the roots of the Hermite polynomial $H_q(x) (j = 1,2,\ldots,q)$, and
the associated weights $ \lambda_j^{(q)}$  are given by
$$\lambda_j^{(q)}= \frac {2^{q-1} q! \sqrt{\pi}} {q^2[H_{q-1}(\xi_j^{(q)})]^2}.$$
As a reference the abscissas (roots) and the weights for $q=6, \; 6$
and $16$ are given in the Appendix. In general, the abscissas and
the weights for the Gauss-Hermite quadrature
 for a given order $q$ can be readily computed, e.g. using the
 functions in \citet{Pres92}.

Applying a change of variable $x=y/\sqrt{2}$ and use the
Gauss-Hermite quadrature to (\ref{eq_intY}), we obtain
  \begin{equation}\label{eq_qy}
\widetilde{Q}_{n-1}(S(t_{n-1}))=\frac{ e^{-r_ndt_n}}{\sqrt{\pi}}
\int_{-\infty}^{+\infty} e^{-x^2} Q_n^{(y)} (\sqrt{2}x) dx \approx
\frac{ e^{-r_n dt_n}}{\sqrt{\pi}} \sum_{j=1}^q \lambda_j^{(q)}
Q_n^{(y)}(\sqrt{2}\xi_j^{(q)}).
\end{equation}
If we apply the change of variable (\ref{eq_y}) and the
Gauss-Hermite quadrature (\ref{eq_qy}) to every grid point $S_m$,
$m=0,1,\ldots,M$, i.e. let $S(t_{n-1})=S_m$, then the option values
at time $t=t_{n-1}$  for all the grid points  can be evaluated
through (\ref{eq_qy}).

\begin{figure}[htbp]
\begin{center}
\includegraphics[scale=0.6]{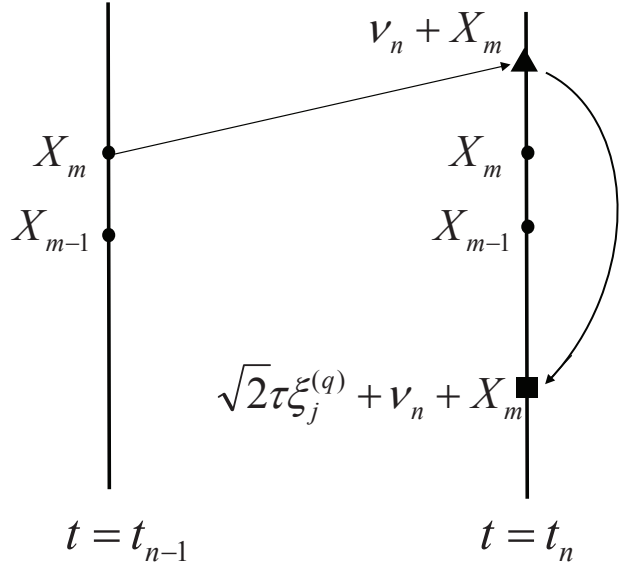}
\vspace{0cm} \caption{Illustration of Gauss-Hermite quadrature
application for an arbitrary grid point $X_m$ at time $t=t_{n-1}$.
The solid circles are fixed grid points, the solid triangle is the
 point of the expected mean at $t=t_n$ given $X_m$ at $t=t_{n-1}$,
 and the solid square is the $j-th$ quadrature point corresponding to $X_m$.} \label{fig_drawing}
\end{center}
\end{figure}

It is a common practice in a finite difference setting for option
pricing to set  the working domain in asset space in
 terms of $X=\ln (S/S(0))$, where $S(0)$ is the spot value at time $t=0$.
 The domain $(X_{\min}, X_{\max})$ is uniformly discretised to yield the grid
 $(X_{\min}=X_0 , X_1=\delta X, X_2=2\delta X,\ldots, X_M=M\delta X=X_{\max})$,
 where $\delta X=(X_{\max}-X_{\min})/M$. The grid point $S_m$, $m=0,1,\ldots,M$, is then given by $S_m=S(0)\exp(X_m)$.
 The boundaries $X_{\min}=\ln(S_{\min}/S(0))$ and
$X_{\max}=\ln(S_{\max}/S(0))$ have to be set sufficiently far
 from spot value  to ensure the adequacy of applying far boundary
conditions.

For each grid point $S_m$ or $X_m$, the variable $Y(t_n)$ is given
by (\ref{eq_y}) with $S(t_{n-1})=S_m$, and the relationship between
$X(t_n)=\ln (S(t_n)/S(0))$ and $Y(t_n)$ for  $S_m$ is worked out to
be $X(t_n)=\tau_n Y(t_n) +\nu_n + X_m$, thus the numerical
integration value for grid point $X_m$ at time $t_{n-1}$ can be
expressed, from (\ref{eq_qy}), as

 \begin{equation}\label{eq_qX}
\widetilde{Q}_{n-1}^{(x)}(X_m) \approx  \frac{ e^{-r_n
dt_n}}{\sqrt{\pi}} \sum_{j=1}^q \lambda_j^{(q)}
Q_{n}^{(x)}(\sqrt{2}\tau \xi_j^{(q)}+\nu_n+X_m).
\end{equation}
where $Q_{n}^{(x)}(X(t_n))$ denotes the option value $Q_{n}(S(t_n))$
as a function of $X(t_n)$ at time $t_n$. The continuous function
$Q_{n}^{(x)}(\cdot)$ is approximated by the cubic spline
interpolation, given the values $Q_{n}^{(x)}(X_m)$ at discrete
points $X_m$, $m=0,1,\ldots,M$. The above description of the
numerical integration using Gauss-Hermite quadrature is illustrated
in Figure \ref{fig_drawing}.

Once the continuation value $\widetilde{Q}_{n-1}^{(x)}$ is known,
then, in  the case of Bermudan option, we get the option price at
time $t=t_{n-1}$ as
  \begin{equation}\label{eq_CX}
Q_{n-1}^{(x)}(X_m)  = \max(\widetilde{Q}_{n-1}^{(x)}\left(X_m) ,
\max(0, \phi\times(S_m-K)\right).
\end{equation}

\subsection{The GHQC algorithm for Bermudan
options}\label{subsec_GHQC}
 The backward time-stepping algorithm
using GHQC for evaluating the expectation (\ref{eq_intS}) and
Bermudan option price (\ref{Bermudan_option_eq}) can be summarized
as follows

\begin{algorithm_new}[GHQC]\label{alg}

~

\begin{itemize}
\item Step 1. Discretize the time domain and asset domain to have time grids $(0 =t_0< t_1 <\cdots<t_N=T)$ and asset grids
(in terms of $X$)  $(X_{\min}=X_0< X_1< \cdots < X_M=X_{\max})$,
where $t_n,\;\;n=1,2,\ldots,N$ are the exercise dates and
$X_m=X_{m-1}+\delta X, \;\;\;\delta X=(X_{\max}-X_{\min})/M,
\;\;m=1,2,\ldots,M$.
\item Step 2. Take final payoff  at maturity $t=t_N=T$ as the option price, i.e. $Q_{N}^{(x)}(X_m)=\max (0, \phi\times(S_m-K)),\;\;m=0,1,\ldots,M$,
where $S_m=S(0)\exp(X_m)$.

\item{Step 3. Do cubic spline interpolation based on variable $X$ and the $M+1$ values  $Q_{N}^{(x)}(X_m),\;\;m=0,1,\ldots,M$. This
is done by first solving the  tri-diagonal system of linear
equations (\ref{eq_tri}), with function values $Q_{N}^{(x)}(X_m)$,
and then use (\ref{eq_g}).

\item Step 4. Do numerical integration for each grid point $X_m$ by Gauss-Hermite quadrature given in (\ref{eq_qX}) to
evaluate $\widetilde{Q}_{N-1}^{(x)}(X_m), \;\;m=0,1,2,\ldots,M$. }
\item Step 5. Apply early exercise test to obtain
$$Q_{N-1}^{(x)}(X_m)  = \max(\widetilde{Q}^{(x)}_{N-1}\left(X_m) , \max(0, \phi\times(S_m-K))\right).$$
\item Step 6. Repeat Steps 3,4 and 5 for time steps $t=t_{N-2}, t_{N-3},\ldots,t_1$.
\item Step 7. Repeat Steps 3 and 4   for time step $t=t_0=0$, and  take  $\widetilde{Q}_0^{(x)}(0)$ as today's option price.

\end{itemize}
\end{algorithm_new}

The above GHQC algorithm has been implemented in $C$ computing
language.

\subsection{GHQC with moment matching}
In calculation of option price expectations (\ref{eq_intY}), the
probability density function (transition density) for $Y(t_n)$ is
known in closed form; it is just standard Normal density. In general
the closed form density function may not be known, and here we
propose a moment matching to replace (\ref{eq_intY}), i.e. assuming
we do not know the density in closed form but we know the moments of
the distribution, we can still use the GHQC algorithm by matching
the numerically integrated moments with the known moments. Let
$p(y)$ denote the unknown probability density function of $Y(t_n)$,
then (\ref{eq_intY}) becomes
\begin{equation}\label{eq_intP}
\widetilde{Q}_{n-1}(S(t_{n-1}))= e^{-r_n dt_n}
\int_{-\infty}^{+\infty}p(y) Q_n^{(y)} (y) dy,
\end{equation}
which can be re-written as
\begin{equation}\label{eq_intP2}
\widetilde{Q}_{n-1}(S(t_{n-1}))= e^{-r_n dt_n}
\int_{-\infty}^{+\infty}e^{-y^2}\times[e^{y^2}p(y) ]Q_n^{(y)} (y)
dy.
\end{equation}
Applying  Gauss-Hermite quadrature (\ref{eq_GHQ}) to
(\ref{eq_intP2}) we then have
 \begin{equation}\label{eq_qG}
\widetilde{Q}_{n-1}(S(t_{n-1})) \approx   e^{-r_n dt_n} \sum_{j=1}^q
\lambda_j^{(q)} \widetilde{p}(\xi_j^{(q)}) Q_n^{(y)}(\xi_j^{(q)}),
\end{equation}
where the function $\widetilde{p}(y)=e^{y^2}p(y)$ is also unknown.
Defining a new weight $W_j^{(q)}=\lambda_j^{(q)}
\widetilde{p}(\xi_j^{(q)}) $, the numerical quadrature for the
integration simplifies to
  \begin{equation}\label{eq_qGW}
\int_{-\infty}^{+\infty}p(y) Q_n^{(y)} (y) dy \approx  \sum_{j=1}^q
W_j^{(q)} Q_n^{(y)}(\xi_j^{(q)}).
\end{equation}
Now we proceed to find the unknown coefficients
$W_j^{(q)},\;j=1,2,\ldots,q$ by matching moments. Recognizing that
if we replace $Q_n^{(y)}(y)$ by $y^K$, the integration yields the
$K$-th moment corresponding to the pdf $p(y)$
 \begin{equation}\label{eq_qM}
\mathrm{E}_{t_{n-1}}[Y(t_n)^K]=\int_{-\infty}^{+\infty}p(y) y^K dy
\approx \sum_{j=1}^q W_j^{(q)}(\xi_j^{(q)})^K.
\end{equation}
If we let $K=0,1,\ldots,q-1$ we then have $q$ equations to determine
the $q$ unknown coefficients $W_j^{(q)},\;j=1,2,\ldots,q$.

In our American option evaluation framework the option value is a
function of $X(t_n)=\ln(S(t_n)/S(0))$, and for each node point $X_m$
we have $X(t_n)=\tau_n Y(t_n) +\nu_n+X_m$. To match the central
moment for random variable $X(t_n)$ (centered at $\nu_n+X_m$),
equation (\ref{eq_qM}) becomes
 \begin{eqnarray}\label{eq_qMX}
\mathrm{E}_{t_{n-1}}[(X(t_n)-\nu_n-X_m)^K]&=&\int_{-\infty}^{+\infty}p_{X(t_n)}(x)
(x-\nu_n-X_m)^K
dx\nonumber\\
&\approx&  \sum_{j=1}^q W_j^{(q)}(\tau_n\xi_j^{(q)})^K,
\;\;K=1,2,\ldots,q,
\end{eqnarray}
where $p_{X(t_n)}(x)$ is the density function of a random variable
$X(t_n)$. For the standard lognormal stock process (\ref{eq_asset}),
the central moments for $X(t_n)$ are simply
$${\text E}_{t_{n-1}}[(X(t_n)-\nu_n-X_m)^K]=
 \left\{\begin{array}{ll}
                    0,\;\; & \text{if } \;K \;\text{ is odd}, \\
                   \tau_n^K(K-1)!!,\;\;&\text{if } \;K \;\text{ is even}, \\
                 \end{array}    \right.$$
where $(K-1)!!$ is the double  factorial, that is, the product of every odd number from $K-1$ to 1.

\begin{remark} Although in (\ref{eq_qMX}) the Gauss-Hermite
weights do not appear explicitly, it is still a direct application
of the full Gauss-Hermite quadrature. To make this clear, we can
substitute back $W_j^{(q)}=\lambda_j^{(q)}
\widetilde{p}(\xi_j^{(q)}) $ in (\ref{eq_qMX}) to obtain a system of
linear equations for the unknown function values
$\widetilde{p}(\xi_j^{(q)}),\;j=0,1,\ldots,q-1 $
 \begin{equation}\label{eq_qMX2}
{\mathrm{E}}_{t_{n-1}}[(X(t_n)-\nu_n-X_m)^K] \approx
\sum_{j=1}^q\lambda_j^{(q)}
\widetilde{p}(\xi_j^{(q)})(\tau_n\xi_j^{(q)})^K, \;\;K=1,2,\ldots,q,
\end{equation}
and obviously solving (\ref{eq_qMX2}) is equivalent to solving
(\ref{eq_qMX}).
\end{remark}


Having found the $q$ coefficients $ W_j^{(q)}$ by solving the system
of linear equations (\ref{eq_qMX}), the expected option value
$\widetilde{Q}_{n-1}^{(x)}(X_m)$ is then approximated as
\begin{equation}\label{eq_qXm}
\widetilde{Q}_{n-1}^{(x)}(X_m) \approx  e^{-r_n dt_n} \sum_{j=1}^q
W_j^{(q)} Q_{n}^{(x)}(\tau_n \xi_j^{(q)}+\nu_n+X_m).
\end{equation}
The GHQC algorithm with moment matching is exactly the same as the one described in the last section, except now
we have to add Step 0 and modify Step 4:

\begin{itemize}
\item Step 0. Find the coefficients  $ W_j^{(q)}$ by solving the system of linear equations (\ref{eq_qMX}).
\item $\ldots$
\item Step 4. Do numerical integration for each grid point $X_m$ by Gauss-Hermite quadrature given in (\ref{eq_qXm}) to
evaluate $\widetilde{Q}_{N-1}^{(x)}(X_m), \;\;m=0,1,\ldots,M$.

\item $\ldots$
\end{itemize}

For convenience we denote the above moment matching algorithm as
GHQC-M.

\subsection{Significant speed up for GHQC}

In the case of lognormal process (\ref{eq_asset}), the number of
time steps required by GHQC for evaluating a Bermudan option
  is the same as the number of exercise dates -- there is no need for using extra  time steps
 between exercise times,  because the numerical integration in GHQC are based on exact transition density of
 the underlying over an arbitrary finite time step. This is true for any discretely monitored Asian, TARN or GMWB annuity contracts.
 On the other hand,    additional time steps
 are often needed by the finite difference method for good accuracy in solving the pde over the finite time step between
  the exercise times.
 The GHQC algorithm described above involves solving a system of linear equations with a tri-diagonal
 matrix for the second derivatives at each time step, which means  GHQC has
  about the same speed per time step as an implicit or semi-implicit finite
 difference  such as the Crank-Nicolson algorithm, which also solves a tri-diagonal system of linear equations at
  each time step. In other words,
 the speed advantage of the original GHQC as described above over finite difference will
 mainly come from using fewer times steps than finite difference.

In numerical practice, however, the GHQC may still need extra time
steps between monitoring times, even though we  have exact
transition density over any finite time steps. This is because for a
larger time step,  more quadrature points  will fall outside the
computational domain. This may be ok for grid points near the
 far boundaries since far boundary conditions can be used for good
approximation for those points, but it will affect the accuracy of
interior grid points if too many quadrature points fall outside the
computational domain,  where extrapolation based on boundary
conditions have to be applied.

The speed advantage of GHQC disappears completely when the
monitoring frequency is high or stochastic process requires fine
time discretization. In such a case it is a model requirement to
discretize time by small steps for small model error. Therefore it
is necessary to make GHQC faster {\it per time step} than finite
difference if we want an overall speed advantage over finite
difference, conditional on
 largely maintaining the accuracy of GHQC.

In the finite difference algorithm with a uniform mesh, the second
spatial derivatives are approximated by the three-point central
difference scheme which has a second order accuracy, while the cubic
spline  is fourth-order in accuracy in the interpolated function
itself. This implies that the cubic spline should correspond to a
second-order accuracy in the second derivatives of the interpolated
function, the same as in a second-order finite difference such as
the Crank-Nicolson algorithm. The above insight gives us a simple
way to speed up the GHQC algorithm significantly. Since the second
derivatives of the interpolated function  have a implied
second-order accuracy, it will not have a material difference if we
use the second-order three-point finite difference to approximate
the second derivatives in the cubic spline formula, thus removing
the need to solve the system of linear equations for the second
derivatives at every time step. In other words, it is perfectly
consistent with the overall accuracy of the cubic spline
interpolation to use the central difference as the second
derivatives, and this simple approximation will significantly speed
up the GHQC algorithm without affecting the overall accuracy of the
algorithm. Indeed, numerical tests show that for the same input of
option values at node points, the simpler cubic spline (using
three-point central difference for the second derivatives) gives
interpolated option values identical in at least the first 9 digits
to those interpolated by the full cubic spline, where the second
derivatives are obtained by solving (\ref{eq_tri}).

 By replacing the second derivatives in the
cubic spline with the three-points central difference, i.e.
$$Q''(x_j)=\left(Q(x_{j+1})+Q(x_{j-1})-2Q(x_j)\right)/\delta x^2$$
and
$$Q''(x_{j+1})=\left(Q(x_{j+2})+Q(x_{j})-2Q(x_{j+1})\right)/\delta x^2,$$
 the solution for the
continuous option value expressed in (\ref{eq_qy}) can now be
calculated by a simple sparse matrix-vector multiplication as
follows
  \begin{equation}\label{eq_C}
{\widetilde{\bm Q}_{n-1}}={\bm H}_n{\bm Q}_n,
\end{equation}
where ${\widetilde{\bm
Q}_{n-1}}=(\widetilde{Q}^{(0)}_{n-1},\ldots,\widetilde{Q}^{(M)}_{n-1})'$
 is the solution vector of size $M+1$ at $t=t_{n-1}^+$, ${\bm
Q}_{n}=({Q}^{(0)}_{n},\ldots,{Q}^{(M)}_{n})'$ is the option value
vector at $t=t_{n}^-$, and ${\bm H}_n$ is a sparse matrix depending
on financial parameters $\mu_n$ and $\sigma_n$, discretization
parameters $dt$ and  $dX$ (uniform nodal spacing) and the numerical
quadratures used. The construction of matrix ${\bm H}_n$ is a simple
exercise of expressing $Q_{n}^{(x)}(\sqrt{2}\tau
\xi_j^{(q)}+\nu_n+X_m)$ in (\ref{eq_qX}) by a linear combination of
$Q_{n}$ values at some neighboring grid points using (\ref{eq_g}),
so for each grid point $X_m$ the value
 $\widetilde{Q}_{n-1}^{(x)}(X_m)$, given by (\ref{eq_qX}),  is after all a simple linear combination of some grid points in ${\bm
Q}_{n}$, i.e. $\widetilde{Q}_{n-1}^{(x)}(X_m) = {\bm H}_n^{(m)} {\bm
Q}_{n}$, where ${\bm H}_n^{(m)}$ is a vector of size $M+1$,
generally sparsely populated by non-zero elements. The global matrix
is then ${\bm H}_n=({\bm H}_n^{(0)},\ldots,{\bm H}_M^{(M)})'$.

For constant financial parameters and equal time steps, this sparse
matrix is fixed and only need to be built once, and each backward
time stepping only involves simply multiplying
 the solution vector at previous time step by a constant sparse
 matrix.
 Now the
GHQC algorithm is fully explicit and it is as fast as an explicit
finite difference, but without suffering the instability inherent in
the explicit finite difference.

Interestingly, after replacing the second derivatives in
(\ref{eq_g}) by the three-point central differences, it can be shown
that the cubic spline
 interpolation (\ref{eq_g}) is equivalent to the following  polynomial interpolation
  through the Lagrange basis polynomials $\{\ell_i\}$
\begin{equation}\label{eq_lag}
Q(x)\approx \sum_{i=j-1}^{j+2} Q(x_i)\ell_i,\;\; \ell_i =
\prod_{j-1\leq m \leq j+2 }^{m\neq i} \frac{x-x_m}{x_i-x_m},\;\; x_j
<x \leq x_{j+1}.
 \end{equation}
The equivalence of (\ref{eq_g}) and (\ref{eq_lag}) can be proved by
some tedious but straightforward algebraic manipulations.
 This equivalence also suggests that higher order Lagrange basis polynomials
 can also be used in our GHQC algorithm to replace the cubic spline
 interpolation, and this higher order interpolation only makes the global matrix ${\bm H}$ slightly
  more densely populated by non-zero entries, but the explicitness of the
 algorithm remains intact.

\section{Numerical Examples}
The current GHQC algorithm is capable of pricing any exotic options
that can be priced by one-dimensional finite difference algorithm.
Here as an illustration of the general accuracy and efficiency of
GHQC, we present  numerical examples of American option pricing and
TARN pricing, comparing GHQC with some of the best performing or
most well-known methods found in the literature, mainly based on the
finite difference method.

In the first set of examples  the American put option has a discrete
exercise feature and the option is called Bermudan option. These
examples were  published in  the original paper
 describing LSMC by \citet{longstaff2001}. The second set of
 examples deal with standard American put options with continuous exercise time,
 published  in \citet{Tangman2008}, which compared some of the
 finest American option pricing algorithms in the literature
 (\citet{Brennan1977},  \citet{Wilmott1995}, \citet{Kwok1997}, \citet{Forsyth2002},
  \citet{Nielsen2002}, \citet{Han2003}, \citet{Ikonen2004} and \citet{Borici2005}). These comparisons are especially meaningful and
 valuable, because the availability of the monotonically convergent
 result of \citet{Leisen1996} as a benchmark or ``true value".
 Some of the above mentioned algorithms are specifically designed for
American options with sophisticated and advanced techniques for
dealing with the free boundary problem of pricing American options.
The current GHQC algorithm does not deal with the free boundary
implicitly - it simply applies the exercise condition explicitly
after each backward time step. At present the GHQC relies on its
high accuracy and high speed per time step to afford very small time
steps to compensate for its lack of sophistication in dealing with
the free boundary in American option pricing. It is a kind of
``brutal force" display.

In the third set of examples we use GHQC to price twelve TARN
options, covering all the three knockout types described in Section
\ref{subsec_TARN} at four different target levels. Results are
compared with those by finite difference and Monte Carlo.

Our GHQC and finite difference algorithms were implemented in C
language, and all our computations were done on a desk PC with
Intel(R) Core(TM) i5-2400 CPU @3.10GHz.

\subsection{Discrete exercise Bermudan puts}
In  the original paper describing LSMC by \citet{longstaff2001}, a
series American options were evaluated by LSMC and
 results were compared with fine solutions of finite difference (FD) method. Here we perform the same computations with the new
 GHQC algorithm and compare results with those of FD and LSMC. For the purpose of comparing both speed and
 accuracy between GHQC and FD, we also implemented a Crank-Nicolson finite difference scheme which has second-order accuracy both in time and space.
 The American exercise constraint in the consistent Crank-Nicolson
 scheme is dealt with by a implicit Projected Successive Over-Relaxation
 (PSOR) scheme (\citet{Wilmott1995}). We have also included our own LSMC calculations for a comparison.
 To estimate the overall accuracy of an algorithm, in this section we
 use the root mean square error of the relative difference (rRMSE)
 between the results of the algorithm in question and the `exact' solution.
Denote $\hat{V}^{(i)},\;i=1,\ldots,m$ as the numerical values of $m$
option prices and $V^{(i)},\;i=1,\ldots,m$ the corresponding true
values, then
$$\text{rRMSE} =\sqrt{\frac{1}{m}\sum_{i=1}^m
\left[\frac{\hat{V}^{(i)}-V^{(i)}}{V^{(i)}}\right]^2}.
$$

The series of test problems reported in  \citet{longstaff2001}
consist of twenty American put options, with
 interest rate fixed at $r=0.06$, drift $\mu=r=0.06$, and strike price fixed at $K=40$. There are five spot prices $S(0)=36, 38,40, 42, 44$,
 two volatilities $\sigma=0.2, 0.4$ and two maturities $T=1, 2$ (years). The combination of those inputs form twenty American
 put options, as listed in Table 1. It is further assumed that the option is exercisable 50 times per year up to
 and including the maturity date $t=T$.

 In  \citet{longstaff2001}, $4\times 10^4$ time steps per year and
1000 steps for the stock price are used for the finite
 difference calculations. In our notation this is $N=4\times 10^4 T$ and $M=1000$. For the purpose of estimating the accuracy
  of LSMC, GHQC and FD (with coarser mesh and larger time steps), those fine solutions of FD could  be regarded as ``exact''
  --
  formally this can be justified
  by a convergence study and error analysis.  We also performed finite difference calculations
  with $N=4\times 10^4 T$ and $M=1000$ and
 we can confirm that the our FD results are identical to
 all the four digits shown in  \citet{longstaff2001}
 for 15 out of the 20 options. The five options for which the two FD  results are not identical in all four digits
 are options with the longer maturity $T=2$ and higher volatility $\sigma=0.4$.
 Excluding these five options, the rRMSE between our FD and those of
 \citet{longstaff2001} with the same fine mesh is $1.13\times 10^{-5}$, while
 separately for the five options the rRMSE is much higher at $5.80\times 10^{-4}$.
 Due to this discrepancy, we did a calculation with doubling of both the
 number of space nodes and time steps, i.e. with $N=8\times 10^4 T$ and $M=2000$.
 Using the  results of this finer mesh as the `exact' values, the rRMSE of  our  FD
 results with  $N=4\times 10^4 T$ and $M=1000$ for all 20 options is
 $3.86\times 10^{-6}$.

 The larger difference between our FD and those of
 \citet{longstaff2001} for the five options with larger volatility and longer
 maturity might suggest the far boundaries in the finite difference
 domain may not be sufficiently far and worth being examined
 carefully. In our implementation (for both FD and GHQC) we have
 set the far boundaries as follows:
\begin{equation}\label{eq_bd}
\begin{array}{l}
S_{\max}=S(0)\exp \left(\max (\nu
T+3\sigma\sqrt{T},3\sigma\sqrt{T})\right), \\
 S_{\min}=S(0)\exp
\left(\min (\nu T-3\sigma\sqrt{T},-3\sigma\sqrt{T})\right),
\end{array}
\end{equation}
where $\nu=\mu-0.5\sigma^2$. The above far boundaries will ensure
that the computation domain always covers at least  three standard
deviations either side of the spot at $t=0$ as well as either side
of the expected mean at maturity $t=T$. Longer maturity and higher
volatility demand wider
 computational domain. Our setting of
 computational domain using (\ref{eq_bd}) automatically responds to both maturity
 and volatility changes.
To make sure these far boundaries are adequate, we did another
calculation with three standard deviations increased to five and
using an even larger number of nodes at $M=3000$. We found that the
rRMSE between solutions with the enlarged boundaries and those with
default boundaries with $N=8\times 10^4 T$ and $M=2000$ is only
$3.49\times 10^{-7}$. So in this study we take our FD solutions with
$N=8\times 10^4 T$ and $M=2000$ as the `exact' solutions for
estimating the relative errors of other algorithms. Table
\ref{Results_table} compares results of GHQC, FD and LSMC. In the
table CN stands for Crank-Nicolson finite difference algorithm
without PSOR, and CN-PSOR stands for CN with PSOR iterations.

\begin{table} [!hbpc]
{\hspace{-1.0cm}{
{\footnotesize{\begin{tabular*}{1.12\textwidth}{ccccccccc} \toprule
 $S(0)$  &  $\sigma$  &   $T$ &  {\bf `Exact' }  & {\bf GHQC } &  {\bf CN-PSOR}  &{\bf  CN}  &  {\bf LSMC} &{\bf  LSMC}$^\ast$\\
 & & & $N=4\times 10^4 T$ & $N=250T$ & $N=1500T$& $N=1500T$& &  \\
 & & & $M=2000$ & $M=200,\;q=5$ & $M=400$& $M=400$& &  \\
 \midrule
36 & 0.2 &1   &4.4778  &4.4779  &4.4780  &4.4777  &4.4811  &4.472 \\
36 & 0.2 &2   &4.8402  &4.8403  &4.8404  &4.8401  &4.8360  &4.821 \\
36 & 0.4 &1   &7.1013  &7.1013  &7.1014  &7.1011  &7.0995  &7.091 \\
36 & 0.4 &2   &8.5068  &8.5065  &8.5068  &8.5066  &8.4829  &8.488 \\
38 & 0.2 &1   &3.2501  &3.2502  &3.2503  &3.2501  &3.2393  &3.244 \\
38 & 0.2 &2   &3.7448  &3.7448  &3.7448  &3.7446  &3.7303  &3.735 \\
38 & 0.4 &1   &6.1476  &6.1476  &6.1476  &6.1474  &6.1358  &6.139 \\
38 & 0.4 &2   &7.6680  &7.6680  &7.6679  &7.6677  &7.6538  &7.669 \\
40 & 0.2 &1   &2.3141  &2.3141  &2.3142  &2.3140  &2.3066  &2.313 \\
40 & 0.2 &2   &2.8846  &2.8845  &2.8846  &2.8844  &2.8725  &2.879 \\
40 & 0.4 &1   &5.3120  &5.3119  &5.3120  &5.3118  &5.3039  &5.308 \\
40 & 0.4 &2   &6.9171  &6.9167  &6.9171  &6.9169  &6.8958  &6.921 \\
42 & 0.2 &1   &1.6170  &1.6170  &1.6170  &1.6169  &1.6112  &1.617 \\
42 & 0.2 &2   &2.2124  &2.2124  &2.2124  &2.2122  &2.2062  &2.206 \\
42 & 0.4 &1   &4.5825  &4.5824  &4.5825  &4.5823  &4.5671  &4.588 \\
42 & 0.4 &2   &6.2443  &6.2443  &6.2442  &6.2440  &6.2316  &6.243 \\
44 & 0.2 &1   &1.1099  &1.1099  &1.1099  &1.1098  &1.1123  &1.118 \\
44 & 0.2 &2   &1.6898  &1.6898  &1.6898  &1.6897  &1.6815  &1.675 \\
44 & 0.4 &1   &3.9477  &3.9477  &3.9477  &3.9475  &3.9388  &3.957 \\
44 & 0.4 &2   &5.6412  &5.6411  &5.6412  &5.6410  &5.6256  &5.622 \\
 \midrule
   &  rRMSE  & & 0.0  &$2.1\times 10^{-5}$  &$2.6\times 10^{-5}$ &$3.4\times 10^{-5}$ &$2.9\times 10^{-3}$  &$3.2\times 10^{-3}$   \\
   &  CPU(sec.) &  & 122 & 0.025  & 8.4  & 0.46 & 20.3 &  \\
\bottomrule
\end{tabular*}
}} \caption{\small{Comparison of Bermudan put option values among
Crank-Nicolson (CN), CN-PSOR, LSMC and GHQC  methods. Strike $K=40$,
interest rate $r=0.06$ and 50 exercise dates per year. Results for
{\bf LSMC}$^\ast$ are taken from \citet{longstaff2001}.}}
\label{Results_table}
}}
\end{table}

In Table \ref{Results_table} for all our results only the first five
digits are shown, while for the LSMC results of
\citet{longstaff2001} only four digits are available. While only
five digits are shown, the calculations of rRMSE have used the full
values without any truncations. For GHQC, we have used a relatively
coarse mesh  ($N=200$) and small number of  time steps ($M=250$,
which is five steps between each exercise dates). The number of
quadrature points is $q=5$. As shown in the table, in terms of rRMSE
the most accurate results belong to GHQC; it has a rRMSE value of
$2.1\times 10^{-5}$ and at the same time it has the least number of
nodes as well as the least number of time steps. While obtaining a
very competitive accuracy, the GHQC has the fastest computing time
by a big margin compared to all the other calculations. For the
finite difference calculations we started with the same  number of
nodes and  number of time steps as the GHQC, and these were
increased gradually until the error in rRMSE more or less matched
that of GHQC. These examples show clearly that significantly larger
number of spatial nodes and time steps are required by FD to match
the accuracy of GHQC. The LSMC estimates are based on $10^5$ (50,000
plus 50,000 antithetic) paths with 50 time steps for each path, the
same number of paths as used in \citet{longstaff2001}. For the basis
functions the first three Laguerre polynomials are used. Due to the
relative slowness of LSMC, we did not attempt to use more
simulations in LSMC to match the accuracy of FD and GHQC in this
study. It is obvious the LSMC is relatively slow and inaccurate in
comparison with FD and GHQC, at least for this set of examples.



The  close values of rRMSE of GHQC (\text{rRMSE}=$2.1\times
10^{-5}$) and FD (\text{rRMSE}=$2.6\times 10^{-5}$) shown in Table
 \ref{Results_table} allows us to do a fair comparison
    of speed of the two algorithms.
    Because the CPU time for each
    option calculation in Table 1 for GHQC and FD is so short, here in Table \ref{Results_table} we
    quote the total CPU time of computing all 20 options. What is
    more, for a more robust estimation of the CPU time, we repeat
    the calculations of all 20 options 100 times and divide the
    total  by 100 to get the  total CPU time for calculating
    all the 20 options once. We found the total CPU time is 20.3 second for LSMC, 8.4 second for FD with PSOR and
    only 0.025 second for GHQC! This is less than 0.0013 second per American put option with an error in
     terms of rRMSE in the order of $10^{-5}$.

    In Table \ref{Results_table} we also show FD calculations
without PSOR, which is much faster than FD with PSOR iterations. In
this case the accuracy of FD without using PSOR is also very good
(\text{rRMSE}=$3.4\times 10^{-5}$),  largely due to the small time
steps used (so that error in explicitly setting the exercise
condition is also small). By not using PSOR the CPU time of FD is
now shortened dramatically from 8 second to 0.46 second. However,
the much shortened CPU time  is still more than 18 times longer than
that of the GHQC for a similar overall accuracy. It is worth
pointing out that a fully explicit  FD can be faster still, perhaps
as fast as GHQC per time step, but our experiments show calculations
with fully explicit FD too often diverged due to its inherent
instability, and when it is convergent the number of time steps is
so large that it is even slower than the semi-implicit
Crank-Nicolson FD without PSOR, and the accuracy of the explicit FD
is also not as good.

The GHQC-M (moment matching alternative) produced results (not
shown) identical to GHQC for at least in the first 10 digits for all
the 20 American put options, and the CPU time is also virtually the
same as GHQC, which is not unexpected, because the extra step for
moment matching in finding the weights in the quadrature only
involves a linear solution of a $q\times q$ matrix.

\subsection{Continuously exercisable American options}
Examples in  the previous section show that for discretely
exercisable American option (i.e. Bermudan option), the GHQC method
requires very few time steps between exercise dates, much fewer than
that required by FD, which is part of the reason why it is so much
faster than FD. In this section, we compare performance of GHQC with
FD and other methods for pricing continuously exercisable American
options, which requires all algorithms to use small time steps to
approximate the continuous exercise feature. Without using
sufficiently small time steps a numerical American price is likely
to contain a material time discretization bias. The requirement for
very small time steps is particularly necessary for the present GHQC
algorithm, because at the moment we do not do anything special about
the free boundary problem, except explicitly setting the exercise
condition after each backward time stepping, while some of the other
methods we are comparing with use some sophisticated techniques to
deal with free boundaries encountered in pricing American options.
In other words, the inherent advantage
 of GHQC not having to use small time steps is totally lost in this
 set of numerical tests.

In \citet{Tangman2008}, numerical results for a set of American put
options with a wide range of financial parameters were compared
among several prominent methods by different authors. In addition to
their own high-order optimal compact finite difference algorithm,
\citet{Tangman2008} included the following algorithms in their
comparison: algorithm by \citet{Brennan1977}, PSOR finite difference
 by \citet{Wilmott1995}, front-fixing finite difference by
 \citet{Kwok1997}, penalty finite difference by \citet{Forsyth2002b},
  penalty and front-fixing method by \citet{Nielsen2002}, transformation finite difference
  algorithm by \citet{Han2003},
  operator splitting algorithm by \citet{Ikonen2004} and the LCP algorithm
 by \citet{Borici2005}. To estimate the accuracy of all the methods,
\citet{Tangman2008} used the option values from the monotonically
convergent binomial method proposed by \citet{Leisen1996} as the
benchmark, or as the `exact' values.

Results for six series of American put options were shown in
\citet{Tangman2008}, three series for $T=0.5$ and three series for
$T=3$, and each series contain  options at five spot values
 and each series corresponds to different combinations of interest rate $r$, dividend $\delta$ and
volatility $\sigma$. According to the results, the series for which
it is most difficult (also most CPU time consuming) to get accurate
solutions is the one with $T=3$ and $\sigma=0.4$, options with the
longest maturity and the highest
 volatility. We computed options in this series with our GHQC and FD (CN-PSOR) algorithms and compare
results with the other algorithms, also using rRMSE against the
`exact' values of the  monotonically convergent binomial method by
\citet{Leisen1996} as the measure of overall accuracy. Table
\ref{Results_table2} shows these results.

\begin{table}[!hbpc]
\hspace{-0.5cm}
{\footnotesize{\begin{tabular*}{1.07\textwidth}{cccccccc} \toprule
method $\backslash$ $S(0)$  &  80   &   90 & 100  & 110 & 120 &  rRMSE  &  CPU (sec.) \\
\midrule
`Exact' & 28.9044 & 24.4482 & 20.7932 & 17.7713 & 15.2560 & N/A                    & N/A \\
GHQC(c)    & 28.9040 & 24.4479 & 20.7930 & 17.7711 & 15.2558 & $1.3\times 10^{-5}$ & 0.022 \\
GHQC(m)    & 28.9043 & 24.4481 & 20.7932 & 17.7713 & 15.2560 & $2.0\times 10^{-6}$ & 0.058 \\
GHQC(f)    & 28.9044 & 24.4482 & 20.7932 & 17.7713 & 15.2560 & $1.1\times 10^{-6}$ & 0.12 \\
CN-PSOR    & 28.9045 & 24.4482 & 20.7932 & 17.7712 & 15.2559 & $3.4\times 10^{-6}$ & 5.42 \\
\citet{Tangman2008} & 28.9045 & 24.4481 & 20.7930 & 17.7708 & 15.2552 & $2.7\times 10^{-5}$ & (0.97) \\
\citet{Brennan1977} & 28.9014 & 24.4422 & 20.7823 & 17.7530 & 15.2271 & $9.9\times 10^{-4}$ & (3.40) \\
\citet{Wilmott1995} & 28.9010 & 24.4416 & 20.7816 & 17.7521 & 15.2259 & $1.0\times 10^{-3}$ & (261)\\
\citet{Borici2005}  & 28.9037 & 24.4463 & 20.7895 & 17.7650 & 15.2458 & $3.5\times 10^{-4}$ & (4.52)\\
\citet{Forsyth2002b} & 28.9012 & 24.4419 & 20.7820 & 17.7527 & 15.2267 & $1.0\times 10^{-3}$ & (8.72) \\
\citet{Nielsen2002} & 28.9088 & 24.4492 & 20.7887 & 17.7587 & 15.2322 & $7.7\times 10^{-4}$ & (10.1) \\
\citet{Ikonen2004}  & 28.9018 & 24.4426 & 20.7827 & 17.7534 & 15.2274 & $9.8\times 10^{-4}$ & (3.97)\\
\citet{Kwok1997}    & 28.9062 & 24.4497 & 20.7951 & 17.7726 & 15.2567 & $6.8\times 10^{-5}$ & (1.41) \\
\citet{Han2003}      & 28.9045 & 24.4479 & 20.7927 & 17.7704 & 15.2548 & $4.3\times 10^{-5}$ & (1.91)\\
 \bottomrule
\end{tabular*}
}} \caption{\small{Comparison of American put option values among
GHQC (with different mesh sizes), FD (CN-PSOR) and methods by other
authors found in the literature. Maturity $T=3.0$, strike $K=100$,
volatility $\sigma=0.4$, interest rate $r=0.07$, dividend
$\delta=0.03$ and spot values $S(0)=(80,90,100,110,120)$. The CPU
times in parentheses are the CPU time quoted by \citet{Tangman2008}
divided by three to compensate for their slower computer with a
1.2GHz clock speed, while our PC has a 3.1GHz clock speed.}}
\label{Results_table2}
\end{table}

For GHQC, we have set the number of quadrature points to $q=16$, the
highest order we have implemented in the code, and used three
different mesh and time step combinations: coarser ($M=300$,
$N=1000T$), median ($M=400$, $N=2000T$) and finer ($M=500$,
$N=3000T$). In Table \ref{Results_table2} the three sets of GHQC
results are denoted by GHQC(c) for the coarser mesh,   GHQC(m) for
the median mesh and GHQC(f) for the finer mesh. For FD (CN-PSOR), we
have used ($M=1000$, $N=6000T$). All the other  option values  in
the table come from the article by \citet{Tangman2008}, which were
obtained using algorithms developed by other authors found in the
literature. Note in \citet{Tangman2008} only the first six digits
were shown for all the option values. Remarkably, our GHQC results
with the finer mesh agree with the `exact' solution in all the first
six digits for all the five American put options. Of course to
estimate the rRMSE error we have used the full un-truncated raw
values. As can be seen from the table, the GHQC algorithm, despite
being the simplest among all the candidates, gives the most accurate
results for pricing this series of American put options and takes
much less time than all the other methods.

Note the CPU times for all the algorithms in Table
\ref{Results_table2} are the CPU time for a single run to obtain all
the five options. That is, after a single backward time stepping
from $t=T$ to $t=0$, the put options at all five different spot
values were obtained at once, interpolating by cubic spline when the
spot value $S(0)$ is not on a grid point. This is true for both our
GHQC and PSOR calculations as well as  all other calculations shown
in Table \ref{Results_table2}. Also, all calculations presented in
\citet{Tangman2008} were done on a computer with 1.2 GHz Intel
Pentium 3 processor (as advised in private communication with the
authors), which is roughly 3 times as slow as our PC with a @3.10GHz
CPU. Therefore for a fair comparison of speed for the algorithms, we
have divided by three the CPU times quoted by \citet{Tangman2008}.
Although this is only a rough conversion from the CPU time on their
computer to the CPU time on our PC, it makes the comparison
reasonably fair. The converted CPU times for all algorithms
presented by \citet{Tangman2008} are shown in parentheses in Table
\ref{Results_table2}. The CPU times for GHQC and FD (CN-PSOR) in our
calculations  is
  obtained by repeating the run 100 times and dividing the total by
 100, to reduce possible influence that can be caused by variations in the CPU clock.

 As shown in Table \ref{Results_table2}, the CPU time of GHQC with the finer mesh, while giving the most accurate results
 at
 rRMSE=$1.1\times 10^{-6}$,  is
 about 8 times as fast as the fastest among all the other
 algorithms.  For the GHQC calculation with the median mesh, the error is at rRMSE=$2.0\times 10^{-6}$ which still significantly smaller than
  all the other algorithms, and the
total CPU
 time  reduces to only 0.058 second. For GHQC with the coarser mesh, the CPU time reduces further to  0.022 second, but  the error,
  at rRMSE=$1.3\times 10^{-5}$,
 is still smaller than  all the other algorithms presented in Table
 \ref{Results_table2} (excluding our own  CN-PSOR),  and at the same time it is 43 times as fast
as  the fastest among all the other algorithms.

In comparison the FD (CN-PSOR) also performed very well but at a
much higher computing cost -- it is more than 43 times slower than
GHQC using the finer mesh, and it is 92 times slower than GHQC using
the median mesh which has a compatible  accuracy to CN-PSOR. Again
the GHQC-M (moment matching alternative) produced results (not shown
in Table \ref{Results_table2}) identical to GHQC for  the first 10
digits for all the 5 American put options, the rRMSE between GHQC
and GHQC-M is in the order of $10^{-10}$, and the CPU time is also
virtually the same as GHQC.

\subsection{Pricing TARN options}
As discussed earlier, for path dependent options such as Asian, TARN
and GMWB, the calculation of expectations between monitoring dates
are the same, there is no special requirement and the GHQC algorithm
can be applied with equal success. The additional requirement is for
applying the jump conditions  across the monitoring dates. Here
there is no practical issue either, because the jump conditions can
be applied exactly the same way as in any finite difference
algorithm. In fact, in GHQC it is more convenient and more natural
to apply the jump conditions than its finite difference counterpart,
because GHQC already engages the full cubic spline interpolation
procedures suitable for accurate interpolation required by the
application of jump conditions.

Specifically, Algorithm \ref{alg} for Bermudan options described in
Section \ref{subsec_GHQC} can be readily modified to price
path-dependent options - Step 5 of applying early exercise is
replaced by applying the proper jump conditions. The application of
jump conditions involves tracking multiple solutions at different
monitoring levels and  interpolating these solutions, exactly the
same tasks performed in a finite difference algorithm. On a high
level, any finite difference algorithm for pricing path-dependent
options, such as Asian, TARN and GMWB,  can be modified to become a
GHQC algorithm by simply replacing the backward time stepping finite
difference  between  monitoring dates with the expectation
calculation using GHQC. In addition, the jump condition application
can be more conveniently performed by GHQC. Here, for a illustration
we compute TARN options using GHQC and compare results with those of
finite difference and Monte Carlo. It should be emphasized that
virtually the same algorithm can be applied to price Asian and GMWB
contracts, the only difference is in the jump conditions. In terms
of numerical algorithms, the difference in jump conditions for
different path dependent options is similar to the difference in
payoff functions in different vanilla options - there is little
effort required to extend the algorithm to other jump conditions
once the algorithm can handle a typical jump condition such as that
encountered in  pricing the TARN options.

\begin{table} [!hbpc]
\begin{center} {

{\footnotesize{\begin{tabular*}{0.8\textwidth}{cccccc} \toprule
 Knockout Type  &  Target  &  {\bf `Exact' }  & {\bf GHQC } &  {\bf FD-CN}  &   {\bf MC} \\
 \midrule
No gain   &0.3 &0.19544 &0.19549 &0.19539 &0.19550 \\
No gain   &0.5 &0.32865 &0.32861 &0.32863 &0.32882 \\
No gain   &0.7 &0.45056 &0.45063 &0.45066 &0.45071 \\
No gain   &0.9 &0.56328 &0.56341 &0.56343 &0.56354 \\
Part gain &0.3 &0.24454 &0.24451 &0.24453 &0.24460 \\
Part gain &0.5 &0.38180 &0.38176 &0.38178 &0.38193 \\
Part gain &0.7 &0.50609 &0.50604 &0.50607 &0.50630 \\
Part gain &0.9 &0.61996 &0.61991 &0.61993 &0.62025 \\
Full gain &0.3 &0.29773 &0.29779 &0.29769 &0.29785 \\
Full gain &0.5 &0.43863 &0.43862 &0.43864 &0.43889 \\
Full gain &0.7 &0.56442 &0.56448 &0.56451 &0.56455 \\
Full gain &0.9 &0.67891 &0.67903 &0.67906 &0.67917 \\
\midrule
      &rRMSE   &N/A     &1.57E-04& 1.49E-04& 4.00E-04\\
      & CPU (Sec)   & 642    &2.18    & 5.25    & 78.1 \\
\bottomrule
\end{tabular*}
}} \caption{\small{Comparison of TARN option values among GHQC,
 finite difference Crank-Nicolson (FD-CN), and Monte Carlo (MC) methods. Spot $S(0)=1.05$, strike
$K=1.0$, domestic and foreign interest rate $r_d=r_f=0$, volatility
$\sigma=0.2$ and 20 payment dates with 30 days between payment
dates. }} \label{table_TARN} }
\end{center}
\end{table}

In the following numerical examples we consider foreign currency
exchange TARN call options with all three types of knockout as
described in Section 2, each knockout type has four cases with four
different targets, so the total number of numerical examples is 12.
The other inputs common to all the examples are spot $S(0)=1.05$,
strike $K=1.0$, volatility $\sigma=0.2$, domestic and foreign
interest rates $r_d=r_f=0$, fixing dates are every 30 days and we
assume 20 fixing dates, which implies the maturity is $T=30\times
20/365\approx 1.6438$. We point out that finite difference solutions
of TARN are rarely found in the literature, and we have implemented
finite difference Crank-Nicolson algorithm for TARN and compared its
performance with that of Monte Carlo (\cite{luo2014pricing}).

Table \ref{table_TARN} compares results among GHQC,
 finite difference Crank-Nicolson (FD-CN), and Monte Carlo (MC)
 methods. In order to estimate the accuracy and efficiency of each
 algorithm with a moderate mesh size (or number of simulations in the case of MC), we have again used finite difference solution from a very fine
 mesh as the `exact' solution, the same as in the previous two
 examples, which can  be actually justified by an error analysis and a convergence
  study. The fine mesh for the finite difference solution has
  $M=2000$, $N=2000$, and the number of grid points for the
  accumulated amount $N_A=500$. Again, the rRMSE of the 12 options is used
  to estimate the accuracy of each algorithm. The CPU time is the
  total time for computing all the 12 options in 12 separate calls
  to the pricing function.

  For GHQC we have set the number of quadrature points $q=6$, and for both GHQC and finite difference (FN-CN) calculations,
  we have  used the same moderately coarse mesh with $M=500$, $N=300$ and
  $N_A=50$. For the Monte Carlo simulations, we have to set the number of simulations to $N_{sim}=1$ million to  achieve an accuracy
   in the same order of magnitude as obtained by GHQC and FD.
 As can be seen from Table  \ref{table_TARN}, with a closely matching
 accuracy, GHQC is more than twice as fast as the finite difference (FD-CN), and both
 GHQC and FD-CN is more accurate than MC with one million
 simulations, and at the same time GHQC is 35 times as fast as Monte
 Carlo, and FD is about 15 times as fast as Monte Carlo.

In addition to  rRMSE given in Table \ref{table_TARN}, the Monte
Carlo also calculates the usual
 standard error of each simulation run. It is interesting to note
 that the average relative standard error (standard error of the mean divided by
 the mean) estimated by Monte Carlo for the 12 simulation
 runs is 4.28E-4, which is quite close to the rRMSE estimate of 4.00E-4 for Monte Carlo
 shown in the table, to some extent justifying the use of a very fine finite difference solution as the `exact' solution
 for estimating numerical errors of GHQC and FD-CN calculations with
 coarser meshes. On the other hand, this close agreement between
 rRMSE and the relative standard error of MC can also be taken as a
 reassuring
  numerical validation of MC simulations for both its mean and error
  estimates.

\section{Conclusions}
 We have  presented a simple, robust and efficient new algorithm for pricing exotic options that can be utilized
  if transition density
 of the underlying asset or its moments are easily evaluated. The new algorithm relies on
 computing the expectations in backward time-stepping through Gauss-Hermite integration quadrature
 applied on a cubic spline interpolation. The essence
 of the new algorithm is the combination of high efficiency and high accuracy numerical integration and interpolation
 on a fixed grid.  It does not  have to be Gauss-Hermite
quadrature for integration and cubic spline for interpolation, other
high order numerical integration and interpolation schemes may
equally be suitable or even superior, depending on the transition
density or moments.
 A `free' bonus of the proposed algorithm is that
it already provides a  procedure for fast and accurate interpolation
of multiple solutions required by many discretely sampled or
monitored path dependent options, such as Asian, TARN and GMWB as
described in the paper. Numerical results of pricing a series of
American options show the accuracy of the new method  can rival many
other very advanced and sophisticated finite difference algorithms,
while at the same time it can be significantly faster than a typical
finite difference scheme. Tests of pricing a series of TARN options
 demonstrate that GHQC is generally capable of pricing path-dependent options with the same  robustness and accuracy
  as the finite difference, but the former can be more efficient. Similar performance is expected for other exotic options such as
barrier, Asian and variable annuities.
For two or three dimensional problems, it remains to be seen if the
new GHQC algorithm has a comparable accuracy and efficiency as the
finite difference method.

\pagebreak

\appendix
\section{Gauss-Hermite quadrature abscissas  and weights}
\begin{table}[htbpc]
\begin{center}
{\footnotesize{\begin{tabular*}{0.5\textwidth}{cc} \toprule
$\xi^{(q)}$  &  $W^{(q)}$  \\
\midrule
\multicolumn{2}{c}{5 point quadrature ($q=5$) } \\
 \midrule
 0.0000000000000000 & 9.4530872048294168E-01 \\
 0.9585724646138185 & 3.9361932315224096E-01 \\
 2.0201828704560856 & 1.9953242059045910E-02 \\
\midrule
\multicolumn{2}{c}{6 point quadrature ($q=6$) } \\
 \midrule
 0.436077411927616 & 7.2462959522439219E-01 \\
 1.33584907401369  & 1.5706732032285659E-01 \\
 2.35060497367449  & 4.5300099055088378E-03 \\
\midrule
 \multicolumn{2}{c}{16 point quadrature ($q=16$) } \\
 \midrule
 0.273481046138152 & 5.0792947901661356E-01\\
 0.822951449144655 & 2.8064745852853262E-01\\
 1.38025853919888  & 8.3810041398985777E-02 \\
 1.95178799091625  & 1.2880311535509970E-02\\
 2.54620215784748  & 9.3228400862418017E-04 \\
 3.17699916197995  & 2.7118600925378804E-05 \\
 3.86944790486012  & 2.3209808448652027E-07\\
 4.68873893930581  & 2.6548074740111637E-10\\
 \bottomrule
\end{tabular*}
}} \end{center}\caption{Gauss-Hermite quadrature abscissas (roots)
and weights for $q=5, \;6$ and  $16$. Only the non-negative
abscissas are given, the negative ones are symmetric about zero with
the same weights.}\nonumber \label{tab_GH}
\end{table}

\pagebreak


\bibliographystyle{chicago} 
\bibliography{bibliography}

\end{document}